\documentclass[12pt]{article}
\pdfoutput=1
\usepackage[utf8]{inputenc}
\usepackage[left=2.55cm, right=2.55cm, top=2.55cm, bottom=2.55cm]{geometry}
\usepackage{tikz}
\usepackage{tikz-feynman}
\usepackage{contour}
\usepackage{amsmath,amssymb,amsbsy}
\usepackage{slashed}
\usepackage{xcolor}
\usepackage{graphicx}
\usepackage{tikz}
\usepackage{tikz-feynman}
\usepackage{contour}
\usepackage{url}
\usepackage{cancel}
\usepackage{cite}
\usepackage[colorlinks=true,allcolors=blue,pdfborder={0 0 0},linktocpage=false,pdfencoding=auto]{hyperref}
\usepackage{tabularx,booktabs}
\usepackage{multicol}
\usepackage{feynmp}
\usepackage{units}
\usepackage{xspace}
\usepackage[labelfont=bf]{caption}
\usepackage[section]{placeins}
\usepackage{subcaption}
\usepackage{soul}
\usepackage{bbold}
\usepackage{parskip}
\usepackage{float}
\usepackage{tabulary}

\definecolor{darkred}{rgb}{0.6,0,0}
\definecolor{darkpurple}{rgb}{0.5,0,0.5}

\newcommand{\beqn}{\begin{eqnarray}}
\newcommand{\eeqn}{\end{eqnarray}}

\begin{document}
\author{
Pran Nath$^a$\footnote{\href{mailto:p.nath@northeastern.edu}{p.nath@northeastern.edu}}\\~\\
$^{a}$\textit{\normalsize Department of Physics, Northeastern University, Boston, MA 02115-5000, USA} \\
}

\title{\vspace{-2cm}\begin{flushright}
\end{flushright}
\vspace{1cm}
\Large \bf Particle physics and cosmology intertwined
 \vspace{0.5cm}}
 
\date{}
\maketitle
\begin{abstract}
While the standard model accurately describes data at the electroweak scale without  
inclusion of gravity, beyond the standard model physics is increasingly intertwined with  gravitational phenomena and cosmology. Thus gravity mediated breaking of supersymmetry in supergravity models lead to sparticles masses, which are gravitational in origin, observable at TeV scales  and testable at the LHC, and supergravity also provides {a} candidate for dark matter, a possible framework for  inflationary models and for models of dark energy. 
 Further, extended supergravity models, and string and D-brane models contain hidden sectors
 some of which may be feebly coupled to the visible sector resulting in heat exchange
 between the visible and hidden sectors. Because of the couplings between the sectors
 both particle physics and cosmology are effected. The above implies that  
  particle physics and cosmology are intrinsically intertwined in the resolution of 
essentially all of the cosmological phenomena such as dark matter and dark energy 
and in the resolution of cosmological puzzles  such as Hubble tension and EDGES anomaly. 
 Here we give a brief overview of the intertwining and implications for the discovery of sparticles, and the resolution of the cosmological anomalies and identification of dark matter and dark energy as major challenges for the coming decades.
  \end{abstract}
\numberwithin{equation}{section}
\newpage

%
\section{Introduction\label{sec1}}
This article is a contribution to Paul Frampton's 80th birthday
   volume marking his over five decades of contributions as a prolific researcher 
   to theoretical physics. He is one of the few theoretical physicists who early on recognized no boundary between particle physics and cosmology and contributed freely  
   to each in good measure. Prominent among his
   works in particle theory relate to physics beyond the standard model and anomaly cancellations
   in higher dimensions, and in cosmology on non-standard cosmological models and  black hole physics.
   Since particle physics and cosmology are the two major areas of his work,  
this paper elaborates on  the progressive intertwining of the fields of particle physics and cosmology,
 over the past several decades from the author's own perspective. 

     For a long period of time up to and including the period of the emergence of the standard model\cite{Glashow:1961tr,Weinberg:1967tq,salam,tHooft:1971qjg,tHooft:1972tcz,Politzer:1973fx,Gross:1973id}         and its tests, it was largely accepted  that gravity could be 
     ignored in phenomena related to particle physics.
     The contrary of course, was not true, as particle
      physics was already known to be central to a variety of astrophysical phenomena such as 
      the Chandrasekhar limit~\cite{Chandrasekhar:1931ih}, and the synthesis of elements
       in the work of B$^2$FH\cite{Burbidge:1957vc} and Peebles \cite{Peebles:1966zz}.      
         For particle physics  gravity became more relevant with the emergence
      of supersymmetry, supergravity and strings. Further, supergravity models in gravity mediated breaking of
      supersymmetry lead to soft terms which allow radiative breaking of the electroweak symmetry
      and predict sparticles observable at colliders. There is another aspect of supergravity and strings
      which has direct impact on particle physics. In extended supergravity, 
      strings and D-brane models one finds hidden sectors which can couple feebly with the visible 
      sector and  affect particle physics phenomena observable at 
    colliders and also have implications for cosmology 
      as they can provide candidates for inflation, dark matter and  dark energy. Thus, with the emergence
      of supergravity and strings deeper connection between particle physics and cosmology has
      emerged. Of course, one hopes that particles physics and cosmology are 
      parts of strings
       and significant literature exists on particle physics-string connection
      (see, e.g., \cite{Dienes:1996du,Nath:2006ut,Ibrahim:2007fb,Abel:2023hkk,Abel:2021tyt}
      and references therein) and on cosmology-string connection (see, e.g.,
      \cite{Kachru:2003aw,Balasubramanian:2005zx,Halverson:2017deq}
      and references therein).      

            In this paper we will focus on the intertwining of particle physics and cosmology. As noted
      above this intertwining has occurred on two fronts: first, in supergravity models with gravity
      mediated breaking   the sparticle
      spectra are direct evidence that gravitational interactions are at work even at the scale of 
      electro-weak physics. {Further, supergravity models with R-parity conservation lead to a candidate
      for dark matter, specifically a neutralino~\cite{Goldberg:1983nd}. Most often
      it turns out to be the lightest supersymmetric particle  in radiative breaking of the electroweak
      symmetry~\cite{Arnowitt:1992aq}. The neutralino as dark matter enters
      importantly  in simulations of cosmological  
      evolution.} At the same time supergravity provides models for the inflationary expansion of the
      universe. Second, also as noted above in extended supergravity and in string models, one finds
      hidden sectors some of which may be feebly coupled to the visible sector. Typically, the hidden sectors and the visible sector will have different temperatures      
      but they have heat exchange which 
      requires a synchronous evolution of the two sectors intertwining the two and affecting
       both particle physics and cosmology.
      The outline of rest of the paper is as follows. In section 2, we will discuss the implications of
      gravity mediated breaking of        
      supergravity at low energy and in section 3 on the intertwining of the particle physics and cosmology
     via hidden sectors.

\section{Gravitational imprint on particle physics at  the electroweak scale}
   As noted above  till the advent of sugra it was the prevalent view that
  gravity did not have much role in particle physics models. 
  However, with the advent of supergravity grand 
  unification\cite{Chamseddine:1982jx,Nath:1983aw}   
  where supersymmetry is broken in the hidden sector
  and communicated to the visible sector by gravitational 
  interactions, one finds that soft breaking terms are dependent 
  on gravitational interactions~\cite{Chamseddine:1982jx,Barbieri:1982eh,Hall:1983iz}.
     Thus, the soft mass of scalars in the
  visible sector $m_s \propto \kappa m^2$, where $\kappa = \sqrt{8\pi G_N}$ and $G_N$ is Newton's constant, and $m$ is an intermediate
  hidden sector mass. Here with $m\sim 10^{10}$ and 
  $M_{Pl}=\kappa^{-1}= 2.43\times 10^{18}$ GeV (in natural units: 
  $\hbar=c=1$), one finds $m_s$ to be of electroweak size. Since  
  sparticle masses are controlled by the soft susy scale, the discovery
  of sparticles would be a signature indicating that gravity has a role in low 
  energy physics.
  This would be very much in the spirit that the discovery of the 
  $W$ and the $Z$ bosons are reflection of $SU(2)_L\times U(1)_Y$
  unification. It is notable that the soft terms are also responsible 
  for generating spontaneous breaking of the electroweak 
  symmetry\cite{Ibanez:1982fr,Chamseddine:1982jx}.  
  An indication that some of the sparticles may be low lying comes
  from the $g-2$ data from Fermilab\cite{Muong-2:2023cdq} which points to a deviation 
  from the standard model prediction of about $4\sigma$.
  An attractive proposition is that the deviation arises from 
  light sparticle exchange, specifically light charginos and 
  light sleptons (see, e.g., \cite{Aboubrahim:2021xfi,Baer:2021aax,Chakraborti:2021dli}  
   and the references therein), a deviation that was predicted 
  quite a while ago\cite{Yuan:1984ww}.  However, a word of 
  caution is in order in that the lattice analysis~\cite{Borsanyi:2020mff} for hadronic vacuum polarization contribution gives a smaller deviation from the standard model than the conventional result where the hadronic vacuum polarization contribution is computed using $e^+e^-\to \pi^+\pi^-$ data. Thus further work is needed in reconciling the lattice analysis with the 
  conventional result on the hadronic polarization contribution before drawing any definitive
  conclusions. 
  
    \section{Hidden sectors intertwine particle physics and cosmology}
As already noted in a variety of  models beyond standard model physics, 
which include extended supergravity models, string models
and extra dimension models, one has hidden sectors. While
these sectors are neutral under the standard model gauge group
they may interact with the visible sectors via feeble interactions.
Such feeble interactions can occur via a variety of portals
which include the Higgs portal\cite{Patt:2006fw}, kinetic energy portal~\cite{Holdom:1985ag,Holdom:1990xp}, Stueckelberg
mass mixing portal\cite{Kors:2004dx,Kors:2004ri}, kinetic-mass-mixing portal\cite{Feldman:2007wj}, Stueckelberg-Higgs portal\cite{Du:2022fqv}, as well as possible
higher dimensional operators. The hidden sectors could be 
endowed with gauge fields, as well as with matter. At the reheat 
temperature the hidden sectors and the visible sector would 
in general lie in different heat baths. However, because of the
feeble interactions between the sectors, there will be heat exchange
between the visible and the hidden sectors and thus their 
thermal evolution will be correlated.  The evolution of the 
relative temperatures of the two sectors then depends on the 
initial conditions, and specifically on the ratio $\xi(T)= T_h/T$
at the reheat temperature,
where $T_h$ is the hidden sector temperature and $T$ is the
visible sector temperature. The Boltzmann equations 
governing the evolution of the visible and the hidden sectors 
are coupled and involve the evolution equation for $\xi(T)$.
Such an equation was derived in 
\cite{Aboubrahim:2020lnr,Aboubrahim:2021dei,Aboubrahim:2022bzk}
 and applied in 
a variety of settings in~\cite{Li:2023nez} consistent with all experimental constraints
on hidden sector matter from terrestrial and astro-physical data~\cite{Aboubrahim:2022qln}.
 It is found that hidden sectors can affect  
observable phenomena in the visible sector, such as the density 
of thermal relics. Hidden sectors provide candidates for dark
matter and dark energy and help resolve cosmological anomalies
intertwining particle physics and cosmological phenomena.
  We discuss some of these topics in further detail below.

Green-Schwarz \cite{Green:1984sg} found that in the low energy limit of Type I strings the kinetic energy of 2-tensor $B_{MN}$  of 10D supergravity multiplet has Yang-Mills and Lorentz group Chern-Simons terms (indicated by superscripts Y and L) so that 
$\partial_{[P} B_{MN]} \to \partial_{[P} B_{MN]}+ \omega_{PMN}^{(Y)}- {\omega_{PMN}^{(L)}}$, where $M,N,P$ are 10-dimensional indices.  Inclusion of the Chern-Simons terms fully requires that one extend the 10D Sugra Lagrangian to  order $O(\kappa)^2$. This was accomplished subsequent to  Green-Schwarz work in \cite{Chamseddine:1986gj} 
 (for related works see \cite{Romans:1985xd,Nishino:1986da,Bergshoeff:1986wc}). 
Dimensional reduction to 4D with a vacuum expectation value for the internal gauge field strength,$\langle F_{ij} \rangle \neq 0$ (where the indices are for the six-dimensional compact
manifold), leads to 
$\partial_{\mu} B_{ij} + A_{\mu} F_{ij} +\, \cdots 
~\sim~ \partial_\mu \sigma + m A_\mu$ ($\mu$ in an index for four dimensional Minkowskian space-time) 
where the internal components $B_{ij}$ give the pseudo-scalar $\sigma$ and $m$ arises from $<F_{ij}>$, which is  a topological quantity, related to the Chern numbers of the gauge bundle. 
Thus $A_\mu$ and $\sigma$ have a Stueckelberg coupling of the form
$A_\mu \partial^\mu \sigma$. 
This provides the inspiration for building BSM models with the 
Stueckelberg mechanism~\cite{Kors:2004dx,Kors:2004ri,Kors:2005uz,Kors:2004iz,Feldman:2006wd}. Specifically, this allows the possibility of 
writing effective theories with gauge invariant mass terms.
For the case of a single $U(1)$ gauge field $A_\mu$ one
may write a gauge invariant mass term by letting $A_\mu\to A_\mu+
\frac{1}{m} \partial_\mu \sigma$ where the gauge transformations are
defined so that $\delta A_\mu= \partial_\mu \lambda$ and $\delta \sigma
= -m\lambda$.  In this case $\sigma$'s role is akin to that of the 
longitudinal component of a massive vector. The above technique
also allows one to generate invariant mass mixing between two 
$U(1)$ gauge fields. Thus consider two gauge groups $U(1)_X$ and
$U(1)_Y$  with gauge fields $A_\mu$, $B_\mu$ and an axionic field
 $\sigma$. In this case we can write a mass term $(m_1 A_\mu
+ m_2 B_\mu + \partial_\mu\sigma)^2$ which is  invariant 
under $\delta_x A_\mu=\partial_\mu \lambda_x$, 
$\delta_x\sigma=-m_1\lambda_x$ for $U(1)_X$, and 
$\delta_y B_\mu=\partial_\mu \lambda_y$, $\delta_y\sigma=-m_2\lambda_y$ for $U(1)_y$. One of the interesting phenomena 
associated with effective gauge theories with gauge invariant mass
terms is that they generate millicharges when coupled to matter fields
\cite{Kors:2004dx,Kors:2005uz,Cheung:2007ut,Feldman:2007wj}.
We will return to this feature of the Stueckelberg mass mixing terms
 when we discuss the EDGES anomaly. 

Hubble tension: Currently there  exists a discrepancy between the measured 
value of the Hubble parameter $H_0$ for low redshifts ($z<1$)
and high redshifts ($z>1000$). Thus for ($z<1$) 
an analysis of data from Cepheids and SNIa gives~\cite{Riess:2021jrx}
$H_0 = (73.04\pm 1.04)\text{ km/s}/\text{Mpc}.$
On the other {hand} an analysis based on $\Lambda CDM$ model 
  the SH0ES Collaboration~\cite{Riess:2021jrx} 
  using 
data from  the Cosmic Microwave Background (CMB), Baryon Acoustic Oscillations (BAO), and from Big Bang Nucleosynthesis (BBN), determines  the Hubble parameter at high $z$  to be~\cite{Planck:2018vyg}
$H_0 = (67.4\pm 0.5)\text{ km/s}/\text{Mpc}.$
This indicates a 5$\sigma$ level tension between the low $z$ and
the high $z$ measurements. 
There is a significant amount of literature attempting to resolve this
puzzle at least partially and recent reviews include 
~\cite{DiValentino:2021izs,Abdalla:2022yfr}.
One simple approach is introducing extra relativistic degrees of 
freedom during the period of recombination which increases the
magnitude of $H_0$ which helps alleviate the tension. 
Models using this idea introduce extra particles such as the
$Z'$ of an extra $U(1)$ gauge field which decays to neutrinos\cite{Gehrlein:2019iwl,Escudero:2019gzq} or utilize 
other particles such as the majoron ~\cite{Fernandez-Martinez:2021ypo,Escudero:2021rfi}.
 The inclusion of extra degrees of freedom, however, 
 must be consistent with the BBN constraints which are 
 sensitive to the addition of massless degrees of freedom. Thus the 
 standard model prediction 
 of $N_{\rm eff}^{\rm SM}\simeq 3.046$~\cite{Mangano:2005cc}
  is consistent with the synthesis of light elements and the
  introduction of new degrees of freedom must maintain 
  {this} successful standard model prediction. The above indicates that the extra
   degrees of freedom should emerge only beyond the BBN time
   and in the time frame of the recombination epoch.
   It is to be noted that new degrees of freedom are also constrained
   by the CMB data as given by the Planck analysis~\cite{Planck:2018vyg}. 
      
A cosmologically consistent model based on the 
 Stueckelberg extension of the SM with a hidden sector
 was proposed in~\cite{Aboubrahim:2022gjb} for alleviating the Hubble tension.
 The model is 
 cosmologically consistent since the analysis is based on 
 a consistent thermal evolution of the visible and the hidden sectors
 taking account of the thermal exchange between the two sectors.
 In addition to the dark fermions, and dark photon, the model
 also contains a  massless pseudo scalar particle field $\phi$
 and a massive long-lived scalar field $s$. The fields $\phi$ and $s$
 have interactions only in the dark sector with no interactions 
 with the standard model fields. The decay of the scalar field occurs
  after BBN close to the recombination time via the decay
  $s\to \phi \phi$ which provides the extra degrees of freedom
  needed to alleviate the Hubble tension. It should be noted that the
  full resolution of the Hubble tension would require going beyond
  providing new degrees of freedom and would involve  
  a fit to all of the CMB date consistent with all cosmological and
  particle physics constraints. For some recent related work on Hubble tension see
\cite{Vagnozzi:2023nrq,Hu:2023jqc,Valdez:2023ros,Zhou:2022wey,Sobotka:2022vrr,Duan:2021whx}.

EDGES anomaly: The 21-cm line plays an important role in the analysis of physics during the dark 
ages and the cosmic dawn in the evolution of the early universe. 
 The 21-cm line arises from the spin transition from the triplet  state to the singlet state
and vice-versa in the ground state of neutral hydrogen. 
The relative abundance of the triplet and the singlet states defines the spin temperature $T_s$ (and $T_B=T_s$)  
of the hydrogen gas and is given by
${n_1}/{n_0} = 3 e^{-{T_*}/{T_s}}$,
where 3 is the ratio of the spin degrees of freedom for the triplet versus the singlet state, $T_*$ is defined
by $\Delta E=k T_*$, where $\Delta E= 1420$ MHz 
is the energy difference at rest between the two spin states, and $T_*\equiv \frac{hc}{k \lambda_{\text{21cm}}}= 0.068 \text{K}$.
 EDGES (The Experiment to Detect the Global Epoch of Reionization Signature) 
reported an absorption profile centered at the frequency $\nu=78$ MHz in the sky-averaged spectrum.The quantity of interest  is the  brightness temperature $T_{21}$
of the 21-cm line defined by
    {$T_{21}(z)=(T_s-T_{\gamma})(1-e^{-\tau})/{(1+z)}$}
   where {$T_{\gamma}(z)$ is the photon temperature at redshift $z$ and}
     $\tau$ is the optical depth for the transition.
The analysis of Bowman et.al\cite{Bowman:2018yin}
finds\footnote{See, however, reference \cite{Hills:2018vyr} on concerns regarding 
 modeling of data.}
 that at  redshift $z\sim 17$,  $T_{21} = - 500_{-500}^{+200}$ mK
at 99\% C.L. On the other hand the analysis of
~\cite{Cohen:2017xpx} based on the $\Lambda$CDM model gives 
 a $T_{21}$ around $-230$ mK,  which shows that the
EDGES result is a 3.8$\sigma$ deviation away from that of the standard cosmological paradigm.

 The EDGES anomaly is not yet confirmed but pending its possible 
 confirmation it is of interest to investigate what possible 
 explanations there might {be}. 
  In fact, several mechanisms have already been 
proposed to explain the 3.8$\sigma$ anomaly~\cite{Munoz:2018pzp,Munoz:2018jwq,Halder:2021jiv,Liu:2019knx,Berlin:2018sjs,Feng:2018rje,Barkana:2018lgd,Barkana:2018cct,Fraser:2018acy,Pospelov:2018kdh,Moroi:2018vci,Fialkov:2019vnb,Choi:2019jwx,Creque-Sarbinowski:2019mcm,Kovetz:2018zan,Bondarenko:2020moh,Lanfranchi:2020crw}. A list of some of the prominent possibilities consist of
  the following: (1) astrophysical phenomena such as radiation from stars and star remnants;
(2)  the CMB background radiation temperature is hotter than expected;
{(3) the} baryons are cooler than what $\Lambda$CDM predicts;
 (4)  modification of cosmological evolution:  inclusion of dark energy such as Chapligin  gas.
 Of the above, there appears to be a leaning towards baryon cooling
 and there is a substantial amount of work in this area following
 the earlier works of \cite{Munoz:2015bca} and 
 Barkana~\cite{Barkana:2018lgd}. Specifically it was pointed out in
 \cite{Barkana:2018lgd} that the observed anomaly could be explained
 if the baryons were cooled down by roughly 3 K.
 Here one assumes a small 
 percentage of DM ($\sim 0.3\%$) is millicharged and  baryons become cooler by Rutherford scattering from the colder  dark matter. As mentioned 
 earlier precisely such a possibility occurs via the Stueckelberg
 mass mixing if we assume one of the gauge fields $U(1)_Y$ is the
 hyper charge gauge field while $U(1)_X$ is a hidden sector field
 and the millicharge dark matter resides in the hidden sector
 while the rest of dark matter could be WIMPS. Within this
 framework a cosmologically consistent analysis of string inspired
 milli-charged model was proposed in\cite{Aboubrahim:2021ohe} 
 where a detailed fit to the data is {possibly} consistent within 
 a high scale model. For some recent work on EDGES anomaly see,
 \cite{Paul:2023emc,Barkana:2022hko,Halder:2022ijw,Wu:2022atm}.

  Inflation:  {As well known} the problems associated with Big Bang 
     such as  flatness, horizon, and the monopole problem  
      are resolved in  inflationary models. In models of this type 
       quantum fluctuations at horizon exit  encode 
 information regarding the characteristics of the inflationary model which can be extracted from the 
  cosmic microwave background (CMB) radiation anisotropy
~\cite{Guth:1980zm,Starobinsky:1980te,Linde:1981mu,Albrecht:1982wi,Linde:1983gd}. In fact data from the Planck experiment
~\cite{Adam:2015rua,Ade:2015lrj,Array:2015xqh} has already
put stringent bounds on inflationary {models} eliminating some.
 {A model proposed in~\cite{Freese:1990rb,Adams:1992bn} is based on
 an axionic field}  with a potential of the form
$V(a) = \Lambda^4 \left(1+ \cos(\frac{a}{f})\right)\,$ 
where $a$ is the axion field and $f$ is the axion decay constant.
However, for the simple model above to hold the Planck data requires
$f> 10M_{Pl}$ which is undesirable since string theory indicates 
that $f$ {lies} below $M_{Pl}$~\cite{Banks:2003sx,Svrcek:2006yi}. However, reduction of $f$ turns out to be a non-trivial issue.
 Techniques used to resolve this issue include the alignment mechanism \cite{Kim:2004rp,Long:2014dta}, n-flation, coherent enhancement \cite{Nath:2019yna}  
  and models using shift symmetry, 
  (for a review and  more references see~\cite{Pajer:2013fsa,Marsh:2015xka}.). \\

 We mention another inflation model which is based in an axion landscape with a $U(1)$ symmetry\cite{Nath:2017ihp}. This model involves $m$ pairs of chiral fields and fields in each pair are oppositely charged under the
  same $U(1)$  symmetry. Our nomenclature is such that we label
  the pseudo-scalar component of each field to be an axion and the
  corresponding real part to be a saxion. Since the model has only 
  $U(1)$ global symmetry, the breaking of the global symmetry 
  leads to just 
 one pseudo-Nambu-Goldstone-boson (PNGB)
   and the remaining pseudo-scalars are not PNGBs.  
Thus the superpotential of the model consists of a part which is
invariant under the $U(1)$ global symmetry and a $U(1)$
symmetry breaking part which simulates instanton effects.
      The analysis of this work shows that the potential contains a fast roll-slow roll splitting mechanism which splits the axion potential into fast roll and slow roll parts where the fields entering fast roll are eliminated
      early on leaving the slow roll part which involves a single axion
      field which drives inflation. Here under the constraints
      of stabilized saxions, one finds inflation models with $f<M_{Pl}$ 
       consistent with Planck data. Similar results are found in the
       Dirac-Born-Infeld based models\cite{Nath:2018xxe}.
       
  Dark energy: One of the most outstanding puzzles of both particle physics and of 
  cosmology is dark energy which constitutes about 70\% of the
energy budget of the universe and is responsible for the accelerated 
  expansion of the universe.  Dark energy is characterized by
  negative pressure so that  $w$ defined by $w=p/\rho$, where
  $p$ is the pressure and $\rho$ the energy density for dark 
  energy, must satisfy $w<-1/3$.  
   { The  CMB and the BAO 
  data fits well} with a cosmological constant $\Lambda$ which corresponds to $w=-1$. Thus the Planck Collaboration~\cite{Planck:2018vyg} gives
  $w=-1.03\pm0.03$ consistent with the cosmological constant. 
    There are two puzzles connected with 
dark energy. First, the use of the cosmological constant
appears artificial, and it is desirable to replace it by a dynamical
field, i.e., a so-called quintessence field (for a review see \cite{Tsujikawa:2013fta}),
which at late times can generate accelerated 
expansion similar to that given by $\Lambda$. The second 
problem relates to the very small size of the cosmological 
constant which is not automatically resolved by simply 
replacing $\Lambda$ by a dynamical field. The extreme fine tuning 
needed in a particle physics model to get to the size of $\Lambda$
 requires a new idea such as vacuum selection in a landscape
with a large number of possible allowed vacua\cite{Weinberg:1988cp}, for instance those
available  in string theory. In any case, it is an example of the
extreme intertwined nature of cosmology and particle physics.
However, finding a quintessence solution that replaces $\Lambda$ and 
is consistent all of the CMB data is itself progress. 
Regarding experimental measurement of $w=-1.03\pm 0.03$ if more accurate
data in future gives
  $w>-1$, it would point to something like quintessence while $w<-1$ would 
  indicate phantom energy and an entirely new sector.

\section{Conclusion} 
  In conclusion it is clear that particle physics and cosmology are deeply intertwined 
  and models of physics beyond the standard model would in the future be 
  increasingly constrained by particle physics experiments as well as by
  astrophysical data. We congratulate Paul for his notable contributions in the
  twin fields and wish him many productive years of contributions for the future.

 {\bf Acknowledgments:} PN is supported in part by the NSF Grant PHY-2209903.   
 

\end{document}